\documentclass[12pt]{iopart}

\usepackage[T1]{fontenc}
\usepackage{verbatim}           
\usepackage{xspace}            
\usepackage{latexsym}           
\usepackage{iopams}             
\usepackage[mathscr]{eucal}     

\newcommand{\GRG}{\textit{Gen. Rel. Grav.}\xspace}
\newcommand{\CMP}{\textit{Comm. Math. Phys.}\xspace}
\newcommand{\APJ}{\textit{Astrophys. J.}\xspace}

\renewcommand{\etal}{\textit{et al}.\@\xspace} 
\newcommand{\viz}{viz.\@\xspace}
\newcommand{\ie}{i.e.\@\xspace}
\newcommand{\eg}{e.g.,\xspace}

\newcommand{\cfr}{cf.\@\xspace}
\newcommand{\dofs}{degrees of freedom\xspace}

\newcommand{\wrt}{with respect to\xspace}
\newcommand{\gr}{general relativity\xspace}
\newcommand{\hog}{higher-order gravity\xspace}
\newcommand{\einspace}{Einstein space\xspace}
\newcommand{\einspaces}{Einstein spaces\xspace}
\newcommand{\rhs}{right-hand side\xspace}

\newcommand{\fn}{\mathfrak{n}}
\newcommand{\Fh}{\mathfrak{H}}
\newcommand{\Fl}{\mathfrak{L}}

\newcommand{\Flb}{\underline{\Fl}}

\newcommand{\cl}{\mathcal{L}}
\newcommand{\cp}{\mathcal{P}}
\newcommand{\cq}{\mathcal{Q}}
\newcommand{\cw}{\mathcal{W}}

\newcommand{\EuD}{\mathscr{D}}      
\newcommand{\EuH}{\mathscr{H}}      
\newcommand{\EuM}{\mathscr{M}}

\newcommand{\rmc}{\mathrm{c}}

\newcommand{\rmz}{\mathrm{z}}
\newcommand{\rmD}{\mathrm{D}}
\newcommand{\rmN}{\mathrm{N}}

\newcommand{\ttT}{\mathtt{T}}

\newcommand{\Lie}{\cl_{\bi{n}}}
\newcommand{\emb}{{^{(3)}} \!}

\newcommand{\nl}{\\ \indent}

\begin{document}

\jl{6}

\title[Bianchi type I spaces in conformal gravity]%
      {Hamiltonian formulation and exact solutions of the Bianchi type I 
       space-time in conformal gravity}

\author{Jacques Demaret%
           \footnote[1]{E-mail: \texttt{demaret@astro.ulg.ac.be}},
        Laurent Querella%
           \footnote[2]{E-mail: \texttt{L.Querella@ulg.ac.be}} and
        Christian Scheen%
           \footnote[3]{E-mail: \texttt{scheen@astro.ulg.ac.be}}}

\address{Institut d'astrophysique et de g\'{e}ophysique de l'universit\'{e} de
         Li\`{e}ge, avenue de Cointe 5, B-4000 Li\`{e}ge, Belgique}

\begin{abstract}
We develop a Hamiltonian formulation of the Bianchi type I space-time in
conformal gravity---the theory described by the Lagrangian 
$\Fl \propto \sqrt{-g} \, C_{abcd} C^{abcd}$, which involves the quadratic 
curvature invariant constructed from the Weyl tensor, in a four-dimensional 
space-time. We derive the explicit forms of the super-Hamiltonian and of the 
constraint expressing the conformal invariance of the theory and we write 
down the system of canonical equations. To seek out exact solutions of this 
system we add extra constraints on the canonical variables and we go through
a global involution algorithm which eventually leads to the closure of the 
constraint algebra. The Painlev\'{e} approach provides us with a proof of 
non-integrability, as a consequence of the presence of movable logarithms in 
the general solution of the problem. We extract all possible particular 
solutions that may be written in closed analytical form. This enables us to 
demonstrate that the global involution algorithm has brought forth the 
complete list of exact solutions that may be written in closed analytical 
form. We discuss the conformal relationship, or absence thereof, of our 
solutions with Einstein spaces.
\end{abstract}

\pacs{9880HW, 0420FY, 0420JB}

\submitted 

\maketitle

\section{Introduction}
\label{sec:intro}

Adopting Weyl's paradigm of a conformally invariant physics \cite{weyl}, 
though without the additional Weyl 1-form, Bach put forth an alternative set 
of fourth-order gravitational field equations---the Bach equations 
\cite{bach}. They are derived by varying, on a four-dimensional Riemannian
space-time $(\EuM,g)$, the unique quadratic conformally invariant action that
is obtained upon gauging the global conformal group \cite{kaku77}, \viz
\begin{equation} \label{ConAct1}
   S = - \frac14 \int_{\EuM} \rmd^4 x \, \sqrt{-g} \, C_{abcd} C^{abcd}, 
\end{equation}
where $C_{abcd}$ is the Weyl tensor.%
\footnote[1]{We adopt Wald's conventions of sign, definitions of curvature 
             tensors (but the extrinsic curvature), and abstract index 
             notation unless otherwise stated \cite{wald}.} 
Bach equations are found in differential geometry and in mathematical studies 
of Einstein's equations of \gr that utilise conformal techniques. For 
instance, Iriondo \etal have recently shown that the integrability conditions 
arising in the null surface formulation of \gr impose a field equation on the
local null surfaces which is equivalent to the fulfilment of the Bach 
equations \cite{irion}. Indeed, the Bach equations constitute a necessary 
condition for a space to be conformal to an \einspace.
\nl
Weyl conformal gravity---the theory based on the action \eref{ConAct1}---has
attracted much interest as a promising candidate for quantum gravity. Beside 
the attractive feature of being the local gauge theory of the conformal group,
quantum conformal gravity possesses the virtues of renormalisability, due to 
the lack of scale in the theory, and asymptotic freedom. Paradoxically,
conformal gravity has not received general acceptance since it involves higher
derivatives; some solutions of the classical theory are expected to have no
lower energy bound and, therefore, exhibit instabilities, \ie `runaway
solutions' \cite{pais}. In other words, ghosts (negative-energy modes)
propagate in the linearised theory, thereby breaking unitarity. However, 
beside the fact that the unitary issue is still open in the context of 
theories with higher derivatives, very interesting results inferred from the 
zero-energy theorem \cite{boulw83} indicate that \emph{nonperturbative} 
effects may alter the spectrum of the asymptotic states, effectively confining
the ghosts \cite{nepom,riege=thesis}. 
\nl
On the other hand, conformal gravity bears no resemblance to \gr classically: 
Einstein gravity is not scale invariant and requires the dimensionful Newton
coupling constant $G_{\rmN}$. Moreover, being fourth-order, the Bach equations
have a larger class of solutions than the second-order Einstein equations; 
those solutions include the aforementioned undesirable runaway modes. 
Notwithstanding this fact, it is a remarkable result that Birkhoff's theorem
holds in conformal gravity, unlike generic quadratic theories 
\cite{riege=birk}. Moreover, every \einspace or every space conformal to an 
\einspace fulfils the vacuum Bach equations automatically. This property has 
led many authors to claim that conformal gravity and Einstein's \gr are 
indistinguishable from each other as far as classical tests are concerned: in 
both theories the Schwarzschild--de Sitter line-element is a vacuum solution.
However, as firstly pointed out by Pechlaner and Sexl \cite{pechl} and 
independently by Havas \cite{havas}, the Schwarzschild--de Sitter metric 
cannot be the vacuum solution of the Bach equations with matter appropriate to 
the exterior of a positive-mass body like the Sun. (Rather than the observed
Newtonian $1/r$ potential, a point-mass $M$ produces a linearly-rising
gravitational potential.) This casts doubts on the viability of Mannheim and
Kazanas's conformal gravity \cite{mannh} (as exemplified by Perlick and Xu's
critics \cite{perli}), which is conceived as a classical alternative to
Einstein's theory and the virtue of which is its ability to account for the
shapes of galactic rotation curves without invoking dark matter.
\nl
Nevertheless, one may consider that conformal gravity provides one with a 
better approximation than \gr to a yet unknown quantum theory of gravity, 
before the expected spontaneous breaking of conformal invariance. In that 
respect, it could make sense to examine whether known properties in 
Einstein's theory also pertain to conformal gravity. With regard to cosmology
this leads us to consider the spatially homogeneous anisotropic spaces of 
Bianchi type I. They are indeed the simplest nontrivial vacuum space-times 
since Friedmann--Lema\^itre--Robertson--Walker (FLRW) isotropic cosmological 
models are conformally flat. In this paper we lay emphasis on the mathematical
analysis rather than the physical discussion of the solutions: Our aim is to
determine whether the (Bianchi type I) Bach equations are integrable or not%
---in a sense that will be made precise later---and to find out exact
solutions. As far as this system is concerned, it turns out that a nice
interplay between involutive methods (within an appropriate canonical 
formalism) and techniques that probe the analytic structure of the system
enables us to go through the integration process and produce the required
solutions. 
\nl
The plan of the paper is as follows. In \Sref{sec:bach} we present briefly the
Bach gravitational field equations and the associated Bach tensor. We also set
our notations for the Bianchi type I metric. We devote \Sref{sec:ham} to the 
canonical formulation and involutive treatment of this system. Two of us have 
already considered the Hamiltonian formulation of Bianchi cosmologies in the
pure $R^2$ variant of the general quadratic theory \cite{demar}. Since in this
paper we consider the conformally invariant case, we build up a Hamiltonian
form of the conformally invariant action \eref{ConAct1} that we further
particularise to the Bianchi type I system. We then take the ensuing canonical
system as our starting point for seeking out exact solutions. For that purpose
we perform the Dirac--Bergmann consistency algorithm on specific constraints
that lead to an involutive system. In \Sref{sec:AnalyticStructure} taking up
the Painlev\'{e} approach we prove the non-integrability of the Bianchi type I
system and extract all possible particular solutions, thereby showing that the 
involution algorithm has produced the complete set of exact solutions that can 
be written in closed analytical form. In \Sref{sec:einspace} we characterise 
our solutions in regard to their conformal relationship with \einspaces. This 
paper is a more thorough presentation of results reported elsewhere 
\cite{quere}.

\section{Conformal gravity and Bach equations}
\label{sec:bach}

Variation of the conformally invariant action \eref{ConAct1} \wrt the metric
yields the conformally invariant fourth-order Bach equations
\begin{equation} \label{Bach1}
   B_{ab} := 2 \nabla^m \nabla^n C_{mabn} + C_{mabn} R^{mn} = 0.
\end{equation}
In equation \eref{Bach1} $B_{ab}$ is called the Bach tensor; it is symmetric,
trace-free, and conformally invariant of weight $-1$. Although expression
\eref{Bach1} has a very compact form this is not appropriate for calculational
purpose with computer algebra: Computation of the contracted double covariant
derivative of the Weyl tensor is quite a heavy task even for simple metrics.
Recently, Tsantilis \etal have provided an algorithm for the 
\textsc{MathTensor} package, which is based on the decomposition of the 
Riemann tensor in its irreducible pieces, and that gives the Bach equations in
a much more tractable form, especially with regard to cosmological 
applications \cite{tsant}. In accordance with their results we write down the 
following equivalent expression of $B_{ab}$, 
\begin{equation} \label{Bach2}
   \fl
   B_{ab} = - \square \Bigl( R_{ab} - \frac{R}6 g_{ab} \Bigr) 
            + \frac13 \nabla_a \nabla_b R
            + \Bigl( C_{mabn} + R_{mabn} + R_{mb} g_{an} \Bigr) R^{mn},
\end{equation}
where the box $\square$ is the d'Alembertian second-order differential 
operator. Note that the equivalence of expressions \eref{Bach1} and 
\eref{Bach2} can be proved owing to the useful formula 
\begin{equation} \label{Love}
   \nabla^m C_{mabc} = \nabla_c L_{ab} - \nabla_b L_{ac},
\end{equation}
where the symmetric tensor $L_{ab}$ is defined by
\begin{equation} \label{LTens}
   L_{ab} = \frac1{12} \Bigl( R g_{ab} - 6 R_{ab} \Bigr),
\end{equation}
and making use of the contracted Bianchi identities.
\nl
The simplest space-times exhibiting nontrivial physical \dofs in the 
conformally invariant gravitational theory based on the action \eref{ConAct1} 
are the spatially homogeneous anisotropic spaces of Bianchi type I---the 
isotropic \textsc{flrw} spaces are conformally flat. The corresponding Bach 
equations \eref{Bach2} can be derived with the help of symbolic computational 
packages such as the \textsc{Excalc} package in \textsc{Reduce} \cite{reduce}. 
The first step is to write the metric in such a way that the conformal 
invariance be manifest \emph{ab initio}. We assume that the Bianchi type I
metric be diagonal. We adopt Misner's parameterisation and choose the conformal
time $\rmd t := \rme^{-\mu} \rmd \tau$ rather than the proper time $\tau$;
hence the line element is 
\begin{equation} \label{Misner}    
   \rmd s^2 = \rme^{2 \mu} \Bigl[ 
                              - \rmd t^2 
                              + \rme^{2 \bigl( 
                                           \beta_+ + \sqrt{3} \beta_- 
                                        \bigr)} \rmd x^2
                              + \rme^{2 \bigl( 
                                           \beta_+ - \sqrt{3} \beta_- 
                                        \bigr)} \rmd y^2
                              + \rme^{- 4 \beta_+} \rmd z^2
                           \Bigr], 
\end{equation}
where $\mu$, $\beta_+$, and $\beta_-$ are functions of time only. The 
independent Bach equations corresponding respectively to the temporal and 
spatial diagonal components of the Bach tensor \eref{Bach2} are given 
explicitly in \ref{app:bacheqn}.

\section{Hamiltonian treatment of Bianchi type I spaces}
\label{sec:ham}

\subsection{Hamiltonian formalism and canonical equations}

Unlike what happens in \gr, where unwanted second-order derivative terms of 
the metric can be discarded from the gravitational action through pure 
divergences, there is no chance whatsoever to weed out such terms in the 
context of \hog theories. It is fortunate however that a consistent method of
building up a Hamiltonian formulation of those theories does actually exist; 
it is a generalisation of the classical \emph{Ostrogradsky formalism} (see, 
\eg \cite{gitma}, and references therein). Basically, it consists in 
introducing auxiliary \dofs that encompass each of the successive derivative 
terms higher than first order. Without resorting explicitly to this method 
Boulware worked out a Hamiltonian formulation of quadratic gravity 
\cite{boulw84}. Contradistinctively, Buchbinder and Lyakhovich developed a 
canonical formalism for the most general quadratic gravitational Lagrangian 
in four dimensions, by employing the aforementioned generalised Ostrogradsky 
method \cite{buchb}. Here we make use of a slightly different generalised 
Ostrograsky construction---as compared to Buchbinder and Lyakhovich's 
formalism---in order to derive a canonical form of the conformally invariant 
action \eref{ConAct1}, which was first obtained by Boulware \cite{boulw84}. 
\nl
Assume that space-time be foliated into a family of Cauchy hypersurfaces
$\Sigma_t$ of unit normal $n^a$ (see, \eg \cite{wald}). The induced metric
$h_{ab}$ onto these hypersurfaces is defined by the formula 
$h_{ab}=g_{ab}+n_a n_b$. The way the hypersurfaces are embedded into space-%
time is provided by the extrinsic curvature tensor 
$K_{ab}:=-h_b^{\ c} \nabla_c n_a = - \case12 \Lie h_{ab}$, 
where $\Lie$ denotes the Lie derivative along the normal $n^a$ (we adopt 
$\Lie$ as a generalised notion of time differentiation). In a coordinate basis
the normal $\bi{n}$ has the components%
\footnote[2]{Latin indices $i,j,k,\dots$ refer to spacelike components in 
             $\Sigma_t$ whereas Greek indices $\alpha, \beta, \gamma, \dots$ 
             refer to components \wrt a specific space-time basis, such as the
             \textsc{adm} basis with vectors  
             $\bi{e}_{\fn} := \bi{n} = (\partial_t - N^i \partial_i)/N$ and 
             $\bi{e}_i = \partial_i$.} 
$n^{\alpha} \equiv (1, - N^i)/N$ and $n_{\alpha} \equiv (-N, 0)$ respectively
and the metric $g_{ab}$ can be written as
\begin{equation}
   \rmd s^2 = h_{ij} \bigl( \rmd x^i + N^i \rmd t \bigr) \otimes
              \bigl( \rmd x^j + N^j \rmd t \bigr) -
              N^2 \rmd t \otimes \rmd t.
\end{equation}
The quantities $N$ and $N^i$ are the standard \textsc{adm} variables, namely
the \emph{lapse function} and \emph{shift vector} respectively. The $3+1$--%
splitting of space-time enables one to express the Lagrangian density
$\Fl= -\case14 \sqrt{-g} \, C_{abcd} C^{abcd}$ only in terms of the quantities
defined onto $\Sigma_t$. For this purpose we make use of the Gauss, Codazzi, 
and York equations, 
\numparts 
\begin{eqnarray} 
   &\emb{R}^a_{\ bcd} = h^a_{\ k} h_b^{\ f} h_c^{\ g} h_d^{\ e} R^k_{\ fge} + 
                        2 K^a_{\ [d} K_{c]b}, \\
   &h_e^{\ d} h_f^{\ c} h_g^{\ b} n_a R^a_{\ bcd} = 2 \rmD_{[f} K_{e]g}, \\
   &n^b n^d h_e^{\ a} h_f^{\ c} R_{abcd} = 
      \Lie K_{ef} + \rmD_{(e} a_{f)} + K_{eg} K^g_{\ f} + a_e a_f,
\end{eqnarray} 
\endnumparts
respectively, where $\rmD$ denotes the covariant derivative onto $\Sigma_t$ 
and $\bi{a}$ is the four-acceleration of an observer moving along the normal 
to $\Sigma_t$, \viz $a_b := n^c \nabla_c n_b$. After some algebra we obtain 
the equations:  
\numparts
\begin{eqnarray} 
\fl h_e^{\ d} h_f^{\ c} h_g^{\ b} n_a C^a_{\ bcd} = 
       2 \Bigl[ 
            h_e^{\ d} h_f^{\ c} h_g^{\ b} - 
            \frac12 h^{bd} \bigl( h_{eg} h_f^{\ c} - h_{fg} h_e^{\ c} \bigr)
         \Bigr] \rmD_{[c} K_{d]b}, \label{Weyl1a} \\
\fl n^b n^d h_e^{\ a} h_f^{\ c} C_{abcd} = 
      \frac12 \Bigl( h_e^{\ a} h_f^{\ c} - \frac13 h_{ef} h^{ac} \Bigr)
              \Bigl( 
                 \Lie K_{ac} + \emb{R}_{ac} + K K_{ac} + 
                 \rmD_{(a} a_{c)} + a_a a_c
              \Bigr). \label{Weyl1b}
\end{eqnarray}
\endnumparts
We can further specify equations \eref{Weyl1a} and \eref{Weyl1b} \wrt the 
\textsc{adm} basis:
\numparts
\begin{eqnarray} 
   C^{\fn}_{\ \, ijk} &= 
      \Bigl[ 
         \delta_i^r \delta_j^s \delta_k^t - 
         \frac12 h^{rt} \bigl( h_{ik} \delta_j^s - h_{ij} \delta_k^s \bigr)
      \Bigr] 
      \Bigl( K_{rs\mid t} - K_{rt\mid s} \Bigr), \label{Weyl2a} \\
   C_{\fn i \fn j} &= 
      \frac12 \Bigl( \delta_i^k \delta_j^l - \frac13 h_{ij} h^{kl} \Bigr)
              \Bigl( 
                 \Lie K_{kl} + \frac{N_{\mid kl}}{N} + K K_{kl} + \emb{R}_{kl}
              \Bigr), \label{Weyl2b}
\end{eqnarray}
\endnumparts
where a stroke `$\mid$' refers to the components of the covariant derivative
in the \textsc{adm} basis and the symbol $\fn$ indicates a component along
$\bi{e}_{\fn}$. Owing to the fact that the Weyl tensor identically vanishes in
three dimensions, the Lagrangian density in \eref{ConAct1} reduces to
\begin{equation} \label{ConAct2}
   \Fl = - N \sqrt{h} \, 
         \Bigl( 
            2 C_{\fn i \fn j} C^{\fn i \fn j} + 
              C_{\fn ijk} C^{\fn ijk} 
         \Bigr),
\end{equation}
where $C_{\fn ijk}$ and $C_{\fn i \fn j}$ are given by equations \eref{Weyl2a}
and \eref{Weyl2b} respectively. 
\nl
Now consider that the induced metric $h_{ij}$ and the extrinsic curvature
$K_{ij}$ be \emph{independent} variables: $K_{ij}$ are introduced as auxiliary
Ostrogradsky variables. In order to recover the definition 
$K_{ij}= - \case12 \Lie h_{ij}$ one we must replace the original Lagrangian 
density by a constrained Lagrangian density,
\begin{equation} \label{ExtLag}
   \Flb = N^{-1} \Fl + \lambda^{ij} \bigl( \Lie h_{ij} + 2 K_{ij} \bigr),
\end{equation}
with Lagrange multipliers $\lambda^{ij}$ as additional variables. One must 
therefore resort to Dirac's formalism for constrained systems (see, \eg 
\cite{henne}, and references therein). The conjugate momenta are defined as 
usual, \viz
\numparts
\begin{eqnarray} 
   p^{ij}   &=& \frac{\partial \Flb}{\partial \Lie h_{ij}}
             =  \lambda^{ij}, \\
   \cq^{ij} &=& \frac{\partial \Flb}{\partial \Lie K_{ij}}
             =  - 2 \sqrt{h} \, C^{\fn i \fn j}, \\
   \Pi^{(\lambda)}_{ij} &=& \frac{\partial \Flb}{\partial \Lie \lambda^{ij}}
                         = 0.
\end{eqnarray}
\endnumparts
(Observe that the variable $\cq^{ij}$ is related to the electric part of the 
Weyl tensor.) Hence there are primary constraints:
\begin{equation} \label{PriCon}
   \phi^{ij} = p^{ij} - \lambda^{ij} \approx 0, \qquad
   \Phi = h_{ij} \cq^{ij} \approx 0, \qquad
   \Pi^{(\lambda)}_{ij} \approx 0.
\end{equation}
(The second constraint arises because only the traceless part of the 
`velocities', $\bigl( \Lie K_{ij} \bigr)^{\ttT}$, can be extracted from 
equation \eref{Weyl2b}.) Performing a Legendre transformation on the extended 
Lagrangian density \eref{ExtLag} we obtain the canonical Hamiltonian density 
\begin{equation} \label{CanHam}
\fl \Fh_{\rmc} = - 2 p^{ij} K_{ij} + \sqrt{h} \, C_{\fn ijk} C^{\fn ijk} 
                 - \frac1{2 \sqrt{h}} \cq^{\ttT ij} \cq^{\ttT}_{ij} 
                 - \cq^{\ttT ij}_{\ \ \ \mid ij} 
                 - \cq^{\ttT ij} \emb{R}_{ij} 
                 - K K^{\ttT}_{ij} \cq^{\ttT ij},
\end{equation}
which is vanishing on account of the fact that any gravitational Lagrangian is
an already parameterised system. Notice that we have discarded a surface
integral at spatial infinity, which reads 
$\oint \rmd \sigma_k 
 \bigl( N_{\mid j} \cq^{jk} - N \cq^{jk}_{\ \ \mid j} \bigr)$. This boundary
term is irrelevant for all class A Bianchi types. We have thus cast the action 
\eref{ConAct2} into canonical form,
\begin{equation} \label{CanAct1}
   S = \int_{\EuM} \rmd^4 x \, N 
          \Bigl[ 
             p^{ij} \, \Lie h_{ij} + \cq^{ij} \, \Lie K_{ij} -
             \Fh_{\rmc} \bigl( h, K, p, \cq \bigr)
          \Bigr].  
\end{equation}
The associated Dirac Hamiltonian density is
\begin{equation} \label{DirHam}
   \Fh_{\EuD} = \Fh_{\rmc} + \mu_{kl} \phi^{kl} + \nu \Phi + 
                \xi^{kl} \Pi^{(\lambda)}_{kl}.
\end{equation}
Consistency of the primary constraints \eref{PriCon} when time evolution is
considered yields the determination of the multipliers $\mu_{kl}$ and 
$\xi^{kl}$ and gives rise to the unique secondary constraint
\begin{equation} \label{SecCon}
   \chi = 2 p + K^{\ttT}_{kl} \cq^{\ttT kl} \approx 0,
\end{equation}
which is first class---it is indeed the generator of conformal 
transformations. Moreover the spurious canonical variables $\cq$ and $K$ can 
be eliminated and only the respective traceless parts of the corresponding 
tensors remain as relevant canonical variables.
\nl
Hitherto the Hamiltonian formalism has been completely general, \ie applicable
to space-times without isometries; henceforth we intend to specify this 
setting to spatially homogeneous cosmologies, that is, Bianchi models, the
metrics of which can be written
\begin{equation} \label{BiaHamMis}    
   \rmd s^2 = - N^2 (t) \rmd t^2 
              + h_{ij}(t) \omega^i \omega^j,
\end{equation}
where $N$ is the lapse function and $\omega^i$ for $i=1,2,3$ are the basis 
dual one-forms (which satisfy 
$\rmd \omega^i = \case12 C^i_{\ kj} \, \omega^j \wedge \omega^k$) that 
characterise the isometry group of each Bianchi type. We adopt the usual
assumptions of Hamiltonian cosmology: we choose the shift vector to be zero 
and assume that the spatial metric $h_{ij}$ be diagonal and parameterised with
Misner's variables $\mu(t)$ and $\beta_{\pm}(t)$. To proceed further we 
perform a canonical transformation that we have defined in a previous work 
\cite{demar} the virtue of which is to disentangle terms stemming respectively 
from the pure $R$-squared and conformal variants of the general quadratic 
theory. The new set of canonical variables is
$\bigl\{ 
\mu, \beta_+, \beta_-; \Pi_{\mu}, \Pi_+, \Pi_-; \cq_+, \cq_-; \cp_+, \cp_- 
 \bigr\}$, 
where $\cq_{\pm}$ and $\cp_{\pm}$ parameterise the traceless tensors
$\cq^{\ttT ij}$ and $K^{\ttT}_{ij}$ respectively; the $\Pi$'s are conjugate to
Misner's variables and the $\cp$'s are conjugate to the $\cq$'s. Hence the
spatial integration in the action \eref{CanAct1} can be performed; the
appropriate canonical action for Bianchi cosmologies is thus given by
\begin{equation} \label{CanAct2}
\fl
   S = \int \rmd t 
          \Bigl[ 
             \Pi_{\mu} \dot{\mu} + \Pi_+ \dot{\beta}_+ + \Pi_- \dot{\beta}_- +
             \cp_+ \dot{\cq}_+ + \cp_- \dot{\cq}_- - 
             N \EuH_{\rmc} - \lambda_{\rmc} \varphi_{\rmc}
          \Bigr],     
\end{equation}   
where the explicit forms of the first-class constraints 
$\EuH_{\rmc} \approx 0$ (super-Hamiltonian) and $\varphi_{\rmc} \approx 0$ 
(conformal constraint) depend on the Bianchi type considered. For Bianchi type
I they reduce respectively to 
\numparts
\begin{eqnarray}
\fl  
   \EuH_{\rmc} = - \frac1{\sqrt{6}} 
                      \Bigl[ 
                         \Pi_+ \cp_+ + \Pi_- \cp_- - 
                         2 \cq_+ \bigl( \cp^2_+ - \cp^2_- \bigr) +
                         4 \cp_+ \cp_- \cq_-     
                      \Bigr]  
           - \frac12 \rme^{- 3 \mu} 
               \Bigl( \cq^2_+ + \cq^2_- \Bigr), \label{SupHam1} \\ 
\fl
   \varphi_{\rmc} = \Pi_{\mu} + \cp_+ \cq_+ + \cp_- \cq_-. \label{ConCon}
\end{eqnarray}
\endnumparts
We need to impose a gauge-fixing condition associated with the conformal
constraint \eref{ConCon}. The introduction of the condition $\mu \approx 0$ as 
an additional constraint has the effect of turning the conformal constraint 
\eref{ConCon} into second class. Hence we are allowed to eliminate the pair of 
canonical variables $\mu$ and $\Pi_{\mu}$. On the other hand, in accordance 
with the form of the Bianchi type I metric \eref{Misner} we can parameterise
the lapse function as $N=\rme^{\mu}$. Therefore, we obtain the final form of 
the canonical action \eref{CanAct2},
\begin{equation} \label{CanAct3}
   S = \int \rmd t \, 
          \Bigl[ 
             \Pi_+ \dot{\beta}_+ + \Pi_- \dot{\beta}_- + 
             \cp_+ \dot{\cq}_+ + \cp_- \dot{\cq}_- - \EuH_{\rmc} 
          \Bigr],     
\end{equation}   
where the super-Hamiltonian \eref{SupHam1} is now given by the expression
\begin{equation} \label{SupHam2}
\fl
   \EuH_{\rmc} = - \frac1{\sqrt{6}} 
                      \Bigl[ 
                         \Pi_+ \cp_+ + \Pi_- \cp_- - 
                         2 \cq_+ \bigl( \cp^2_+ - \cp^2_- \bigr) +
                         4 \cp_+ \cp_- \cq_-    
                      \Bigr]
                 - \frac12 \Bigl( \cq^2_+ + \cq^2_- \Bigr).  
\end{equation}
\indent
Varying the action \eref{CanAct3} with respect to the remaining canonical 
variables and their conjugate momenta we derive the canonical equations of
conformal gravity: 
\numparts
\begin{eqnarray}
   \dot{\beta}_{\pm} &=& - \frac1{\sqrt{6}} \cp_{\pm}, \label{CanEq1a} \\ 
   \dot{\Pi}_{\pm}   &=& 0, \label{CanEq1b} \\
   \dot{\cq}_+       &=& - \frac1{\sqrt{6}} 
                           \bigl( 
                              \Pi_+ + 4 \cp_- \cq_- - 4 \cp_+ \cq_+ 
                           \bigr), 
                           \label{CanEq1c} \\
   \dot{\cq}_-       &=& - \frac1{\sqrt{6}} 
                           \bigl( 
                              \Pi_- + 4 \cp_- \cq_+ + 4 \cp_+ \cq_- 
                           \bigr), 
                           \label{CanEq1d} \\
   \dot{\cp}_+       &=&   \frac2{\sqrt{6}} \bigl( \cp^2_- - \cp^2_+ \bigr) + 
                           \cq_+, \label{CanEq1e} \\         
   \dot{\cp}_-       &=&   \frac4{\sqrt{6}} \cp_+ \cp_-  + \cq_-. 
                           \label{CanEq1f}         
\end{eqnarray} 
\endnumparts
This system supplemented with the super-Hamiltonian constraint \eref{SupHam2} 
is equivalent to the system of fourth-order field equations \eref{BachEq00}--%
\eref{BachEq33}. However, instead of the Bach equations we have now at our 
disposal the nice differential system \eref{CanEq1a}--\eref{CanEq1f} and the
algebraic constraint \eref{SupHam2}. We shall seek out exact solutions by
performing a global involution algorithm on appropriate extra constraints that
yield closed constraint algebras. 

\subsection{Global involution of extra constraints}
\label{subsec:GloInv}

The involution method consists in applying the Dirac--Bergmann consistency
algorithm to our classical system, with the Poisson bracket defined \wrt the 
canonical variables $\beta_{\pm}$, $\Pi_{\pm}$, $\cq_{\pm}$, $\cp_{\pm}$, and 
after suitable conditions have been imposed. For a rigorous treatment on the 
concept of involution as applied to constrained systems, we refer the 
interested reader to Seiler and Tucker's article \cite{seile}. Strictly 
speaking, the steps of the global involution algorithm are:
\begin{enumerate}
   \item Impose an appropriate extra constraint on the canonical variables.
         \label{step1}
   \item Require that constraint to be preserved when time evolution is
         considered. This gives rise to secondary constraints and possibly to
         the determination of the Lagrange multiplier associated with the
         extra constraint. \label{step2}
   \item Repeat Step (\ref{step2}) until no new information comes out. 
         \label{step3}
\end{enumerate}
Once the involution algorithm has been performed we can classify all the
constraints into first class and second class and proceed further to the
analysis of the particular system. 
 
\subsubsection{Constraint $\varphi_p \approx 0$.}
\label{subsubsec:Phip}

The first extra constraint we consider expresses that the ratio of the 
variables $\cp_{\pm}$ be constant, namely
\begin{equation} \label{Phip0} 
   \varphi_p^{(0)} = \cp_- - p \cp_+ \approx 0,  \qquad p \in \mathbb{R}. 
\end{equation}   
The Poisson bracket of $\varphi_p^{(0)}$ and $\EuH_{\rmc}$ yields the secondary
constraint
\begin{equation} \label{Phip1}
   \varphi_p^{(1)} = \frac2{\sqrt{6}} p \bigl( p^2 - 3 \bigr) \cp^2_+ -
                     \bigl( \cq_- - p \cq_+ \bigr) \approx 0. 
\end{equation}   
The Poisson bracket of $\varphi_p^{(1)}$ and 
$\EuH_{\EuD} := \EuH_{\rmc} + \lambda_p \varphi_p^{(0)}$ leads to the 
determination of the Lagrange multiplier, \viz
$\lambda_p \approx (\Pi_- - p \Pi_+)/[\sqrt{6} (1 + p^2)]$,  
which turns out to be constant, and the algorithm stops. Both constraints 
\eref{Phip0} and \eref{Phip1} are second class since their Poisson bracket 
is equal to $1 + p^2$. Thus we can eliminate the corresponding spurious \dofs. 
To this end we perform the canonical transformation 
\begin{equation} \label{CanTra}
   \cq_{\pm} \rightarrow \cq_{\pm} \pm p \cq_{\mp}. 
\end{equation}   
The action \eref{CanAct3} then reduces to 
\begin{equation} \label{CanAct4}
   S = \int \rmd t \, 
          \Bigl[ \Pi_+ \dot{\beta}_+ + 
                 \Pi_- \dot{\beta}_- + 
                 \cp_+ \dot{\cq}_+ - \EuH_{\rmc} 
          \Bigr],     
\end{equation}   
where the super-Hamiltonian \eref{SupHam2} is now given by
\begin{equation} \label{SupHam3}
\fl
   \EuH_{\rmc} = - \frac1{\sqrt{6}} \bigl( \Pi_+ + p \Pi_- \bigr) \cp_+ 
                 - \frac2{\sqrt{6}} 
                      \Biggl( \frac{3 p^2 - 1}{1 + p^2} \Biggr) 
                      \cp^2_+ \cq_+  
                 + \Biggl[ 
                      \frac{p^2 \bigl( p^2 - 3 \bigr)^2}
                           {3 \bigl( 1 + p^2 \bigr)} 
                   \Biggr] \cp^4_+ 
                 - \frac{\cq^2_+}{2 \bigl( 1 + p^2 \bigr)}.
\end{equation}
Varying the action \eref{CanAct4} with respect to the pair of canonical 
variables $\cq_+$ and $\cp_+$ we obtain the canonical equations
\numparts
\begin{eqnarray}
\fl
   \dot{\cq}_+ = - \frac1{\sqrt{6}} \bigl( \Pi_+ + p \Pi_- \bigr)
                 - \frac4{\sqrt{6}} 
                   \Biggl( \frac{3 p^2 - 1}{1 + p^2} \Biggr) \cp_+ \cq_+
                 + \Biggl[
                      \frac{4 p^2 \bigl( p^2 - 3 \bigr)^2}
                           {3 \bigl( 1 + p^2 \bigr)^2}
                   \Biggr] \cp^3_+, 
                 \label{CanEq2a} \\
\fl
   \dot{\cp}_+ =   \frac2{\sqrt{6}} 
                   \Biggl( \frac{3 p^2 - 1}{1 + p^2} \Biggr) \cp^2_+ 
                 + \frac{\cq_+}{1 + p^2}. \label{CanEq2b}         
\end{eqnarray} 
\endnumparts
Taking into account the explicit form of the super-Hamiltonian \eref{SupHam3}
we obtain a nonlinear first-order differential equation for $\cp_+ (t)$,
\begin{equation} \label{BriBou0}
   \dot{\cp}^2_+ = \frac23 \bigl( 1 + p^2 \bigr) \cp^4_+ -
                   \frac2{\sqrt{6}} 
                   \Biggl( \frac{\Pi_+ + p \Pi_-}{1 + p^2} \Biggr) \cp_+,  
\end{equation}   
which can be written as one binomial equation of Briot and Bouquet, 
\begin{equation}
   \dot{u}^2 = u^4 + \gamma^3 u,
\end{equation}   
in terms of a new variable $u$ defined by the homography
$u = \pm 2 \cp_+ \sqrt{\bigl( 1 + p^2 \bigr)/6}$,
with 
$\gamma^3 = \mp \case23 \bigl( \Pi_+ + p \Pi_- \bigr)
 \bigl( 1 + p^2 \bigr)^{-1/2}$.
Direct calculation shows that its general solution is given by 
\begin{equation} \label{BriBou1}
   u_1 (t) = \frac{\gamma^3}{4 \cw}, 
\end{equation}   
where $\cw$ stands for the Weierstrass elliptic function 
$\wp (t - t_0; g_2, g_3)$, with invariants $g_2 = 0$ and 
$g_3 = - \case1{36} \bigl( \Pi_+ + p \Pi_- \bigr)^2/(1 + p^2)$, and with one 
arbitrary constant $t_0$ (see \cite{abram} pp~627~ff). In terms of the 
canonical pair $(\cq_+, \cp_+)$ the representation \eref{BriBou1} corresponds 
to the expressions 
\begin{equation} \label{BBSol}
\eqalign{%
\fl
   \cq_+ (t) &= \frac{\sqrt{6}}{72} 
                \frac{\bigl( \Pi_+ + p \Pi_- \bigr)}{\cw^2} 
                   \Biggl\{ 
                      \Biggl[ 
                         \frac{1 - 3 p^2}{\bigl( 1 + p^2 \bigr)^2} 
                      \Biggr]
                      \bigl( \Pi_+ + p \Pi_- \bigr) + 
                      6 \cw \frac{\rmd \ln \cw}{\rmd t}
                   \Biggr\}, \\  
\fl
   \cp_+ (t) &= - \frac{\sqrt{6}}{12}  
                  \Biggl( \frac{\Pi_+ + p \Pi_-}{1 + p^2} \Biggr) 
                  \frac1{\cw},} 
\end{equation}   
where                           
\begin{equation*}                          
   \cw \frac{\rmd \ln \cw}{\rmd t} = 
       \pm \frac16 \sqrt{144 \cw^3 + 
                         \frac{\bigl( \Pi_+ + p \Pi_- \bigr)^2}{1 + p^2}}. 
\end{equation*}   
To obtain the analytic form of the metric functions $\beta_{\pm} (t)$ we must 
integrate the \rhs of the expression giving $\cp_+ (t)$. We apply the
following theorem (see \cite{abram} p~641): If $\wp^{\prime} (z_0) \neq 0$,
then 
\begin{equation*}
   \wp^{\prime} (z_0) \int \frac{\rmd z}{\wp (z) - \wp (z_0)} 
      = 2 z \zeta (z_0) + \ln \sigma (z-z_0) - \ln \sigma (z+z_0),    
\end{equation*}   
with Weierstrassian functions $\zeta (z)$ and $\sigma (z)$ defined by 
$\zeta^{\prime} (z) + \wp (z) = 0$ and 
$\sigma^{\prime} (z) - \sigma (z) \zeta (z) = 0$ respectively. To achieve our 
aim we take $z=t$ and $z_0=t_{\rmz}$, where $t_{\rmz}$ is a zero of the
Weierstrass elliptic function, \ie $\wp (t_{\rmz}) = 0$. Indeed, we obtain
\begin{equation} \label{IntWeier}
   \int \frac{\rmd t}{\cw (t)} 
      = \pm \Biggl[ \frac{6 \sqrt{1 + p^2}}{\Pi_+ + p \Pi_-} \Biggr]
        \Bigl[ 
           2 t \zeta (t_{\rmz}) + 
           \ln \sigma (t-t_{\rmz}) - \ln \sigma (t+t_{\rmz})
        \Bigr],    
\end{equation}   
and we can write the corresponding homogeneous metrics of type I under the 
form 
\begin{equation} \label{MetPhip}
\eqalign{%
   \rmd s^2 = - \rmd t^2  
              &+ \exp \Biggl[ 
                         \pm \frac{2(1+\sqrt{3}p)}
                                  {\sqrt{1+p^2}} t \zeta(t_{\rmz})
                      \Biggr]
                 \Biggl[ 
                    \frac{\sigma (t-t_{\rmz})}{\sigma (t+t_{\rmz})}
                 \Biggr]^{\pm \frac{1+\sqrt{3}p}{\sqrt{1+p^2}}} \rmd x^2 \\
              &+ \exp \Biggl[ 
                         \pm \frac{2(1-\sqrt{3}p)}
                                  {\sqrt{1+p^2}} t \zeta(t_{\rmz})
                      \Biggr]
                 \Biggl[ 
                    \frac{\sigma (t-t_{\rmz})}{\sigma (t+t_{\rmz})}
                 \Biggr]^{\pm \frac{1-\sqrt{3}p}{\sqrt{1+p^2}}} \rmd y^2 \\
              &+ \exp \Biggl[ 
                         \mp \frac4{\sqrt{1+p^2}} t \zeta(t_{\rmz})
                      \Biggr]
                 \Biggl[ 
                    \frac{\sigma (t-t_{\rmz})}{\sigma (t+t_{\rmz})}
                 \Biggr]^{\mp \frac2{\sqrt{1+p^2}}} \rmd z^2.} 
\end{equation}

\indent
As a particular case of the present analysis we can specialise the above
solution \eref{BriBou1} to the axisymmetric case, for which the secondary 
constraint \eref{Phip1} reduces to the conditions
\begin{equation} \label{AxiCon}
   \frac{\cp_-}{\cp_+} \approx p \approx \frac{\cq_-}{\cq_+},
   \qquad p \in \bigl\{ 0, \pm \sqrt{3} \bigr\}.
\end{equation}   
Owing to the canonical transformation \eref{CanTra} the new variable $\cq_-$ 
vanishes automatically. Hence the super-Hamiltonian \eref{SupHam3} becomes 
\begin{equation} \label{SupHam4}
   \EuH_{\rmc} = - \frac1{\sqrt{6}} \bigl( \Pi_+ + p \Pi_- \bigr) \cp_+ 
                 - \frac2{\sqrt{6}} 
                      \Biggl( \frac{3 p^2 - 1}{1 + p^2} \Biggr) 
                      \cp^2_+ \cq_+  
                 - \frac{\cq^2_+}{2 \bigl( 1 + p^2 \bigr)}
\end{equation}
and the corresponding canonical equations are
\numparts
\begin{eqnarray}
   \dot{\cq}_+ &=& - \frac1{\sqrt{6}} \bigl( \Pi_+ + p \Pi_- \bigr)
                   - \frac4{\sqrt{6}} 
                     \Biggl( \frac{3 p^2 - 1}{1 + p^2} \Biggr) \cp_+ \cq_+,
                   \label{CanEq3a} \\
   \dot{\cp}_+ &=&   \frac2{\sqrt{6}} 
                     \Biggl( \frac{3 p^2 - 1}{1 + p^2} \Biggr) \cp^2_+ 
                   + \frac{\cq_+}{1 + p^2}. \label{CanEq3b}         
\end{eqnarray} 
\endnumparts
As in the general case we find out the binomial equation of Briot and Bouquet 
\eref{BriBou0} the solution \eref{BriBou1} of which is valid for any value of 
the parameter $p$. Hence the axisymmetric solution is obtained by setting $p$ 
to $0$ or $\pm \sqrt{3}$ in equation \eref{BBSol} and finally in the metric 
\eref{MetPhip}. 

\subsubsection{Constraint $\varphi_q \approx 0$.}
\label{subsubsec:Phiq}

Consider the extra constraint expressing that the ratio of the variables 
$\cq_{\pm}$ be constant, namely
\begin{equation} \label{Phiq0}
   \varphi_q^{(0)} = \cq_- - q \cq_+ \approx 0,  \qquad q \in \mathbb{R}. 
\end{equation}   
The Poisson bracket of $\varphi_q^{(0)}$ and 
$\EuH_{\EuD} := \EuH_{\rmc} + \lambda_q \varphi_q^{(0)}$ yields the secondary 
constraint
\begin{equation} \label{Phiq1}
   \varphi_q^{(1)} =  4 \cq_+ \Bigl[ 
                                 \cp_- \bigl( q^2 - 1 \bigr) - 2 q \cp_+
                              \Bigr] -
                      \bigl( \Pi_- - q \Pi_+ \bigr) \approx 0. 
\end{equation}   
To check the consistency of the involution algorithm we consider two distinct 
cases: 
\begin{enumerate}
   \item The secondary constraint \eref{Phiq1} is identically satisfied if 
         $\Pi_- - q \Pi_+ \approx 0$ and 
         $\cp_- (q^2 - 1) - 2 q \cp_+ \approx 0$. Again, in order to proceed, 
         we split up the analysis into two subdivisions: 
         \begin{enumerate}
            \item If $q \neq \sigma$, with $\sigma=\pm 1$, the following weak 
                  equality holds: $\cp_- \approx 2 q \cp_+ / (q^2 - 1)$. The 
                  Poisson bracket of $\varphi_q^{(1)}$ and $\EuH_{\EuD}$ 
                  yields the constraint
                  \begin{equation} 
                  \fl
                     \varphi_q^{(2)} = \bigl( 1 - 3 q^2 \bigr) \lambda_q +
                                       q \bigl( q^2 - 3 \bigr) \cq_+ -
                                       \frac{4 q \bigl( 1 - 3 q^2 \bigr)
                                                 \bigl( q^2 - 3 \bigr)}
                                            {\sqrt{6} \bigl( q^2 - 1 \bigr)^2}
                                       \cp^2_+ \approx 0, 
                  \end{equation}   
                  where $q \neq \sigma/ \sqrt{3}$, for we consider nonzero 
                  canonical variables $\cq_{\pm}$. Consistency then leads to 
                  the determination of the Lagrange multiplier $\lambda_q$, 
                  \viz
                  \begin{equation*}
                     \lambda_q \approx 
                        \Biggl[ 
                           \frac{q \bigl( q^2 - 3 \bigr)}{3 q^2 - 1} 
                        \Biggr] \cq_+ +
                        \Biggl[ 
                           \frac{4 q \bigl( q^2 - 3 \bigr)}
                                {\sqrt{6} \bigl( q^2 - 1 \bigr)^2}
                        \Biggr] \cp^2_+.
                  \end{equation*}   
                  When $q \in \bigl\{ 0, \sigma \sqrt{3} \bigr\}$ the Poisson 
                  bracket of $\varphi_q^{(2)}$ and $\EuH_{\EuD}$ vanishes 
                  identically and the involution algorithm does not generate 
                  any new constraint. This subcase corresponds to a particular
                  case of the axisymmetric solution with 
                  $\Pi_- \approx q \Pi_+$ (\cfr equation \eref{AxiCon}). For 
                  the nonaxisymmetric case, on the other hand, the next step 
                  in the involution algorithm yields 
                  \begin{equation*}
                     \Pi_+ \approx 
                        \Biggl[ 
                           \frac{4 \bigl( q^2 + 1 \bigr)}{q^2 - 1} \cq_+ + 
                           \frac{16 \bigl( 3 q^2 - 1 \bigr) 
                                    \bigl( q^2 + 1 \bigr)}
                                {\sqrt{6} \bigl( q + 1 \bigr)^3 
                                          \bigl( q - 1 \bigr)^3} \cp^2_+
                        \Biggr] \cp_+,
                  \end{equation*}   
                  and the last step gives
                  \begin{equation*}
                     \cq_+ \approx \frac{\bigl( 3 q^2 - 1 \bigr) 
                                         \bigl( \sigma - 3 \bigr)}
                                        {\sqrt{6} \bigl( q^2 - 1 \bigr)^2} 
                     \cp^2_+. 
                  \end{equation*}   
                  At this stage no more constraints arise. Taking into account 
                  that the super-Hamiltonian \eref{SupHam2} is weakly
                  vanishing we find out that $\Pi_+$ must be equal to zero. 
                  This subcase will be discussed in \S~\ref{subsubsec:PiCons},
                  as part of our discussion of the generic case with 
                  $\Pi_\pm \approx 0$.
            \item If $q = \sigma$, then $\cp_+$ is weakly vanishing. The 
                  Poisson bracket of $\varphi_q^{(1)}$ and $\EuH_{\EuD}$ 
                  yields the constraint 
                  \begin{equation} 
                     \varphi_q^{(2)} = \lambda_q + \sigma \cq_+ + 
                                       \frac{2 \sigma}{\sqrt{6}} \cp^2_- 
                                       \approx 0 
                  \end{equation}   
                  and consistency determines the Lagrange multiplier 
                  $\lambda_q$. The Poisson bracket of $\varphi_q^{(2)}$ and 
                  $\EuH_{\EuD}$ gives the weak equality
                  \begin{equation*}
                     \Pi_+ \approx 
                        \Bigl( 
                           \frac4{\sigma} \cq_+ + 
                           \frac8{\sigma \sqrt{6}} \cp^2_-
                        \Bigr) \cp_-.
                  \end{equation*}   
                  The last step in the involution algorithm yields
                  \begin{equation*}
                     \cq_+ \approx - \frac{3 + \sigma}{2 \sqrt{6}} \cp^2_-. 
                  \end{equation*}   
                  At this stage no more constraints arise. The vanishing of 
                  the super-Hamiltonian \eref{SupHam2} restricts $\Pi_+$ to be
                  zero. As in the previous nonaxisymmetric case with 
                  $q \neq \sigma$, this subcase will be discussed in 
                  \S~\ref{subsubsec:PiCons}.
         \end{enumerate}
   \item When $\Pi_- - q \Pi_+ \neq 0$, the Poisson bracket of 
         $\varphi_q^{(1)}$ and $\EuH_{\EuD}$ yields the constraint 
         \begin{eqnarray*} 
         \fl
            \varphi_q^{(2)} = 
               \sqrt{6} q \bigl( q^2 - 3 \bigr) \cq^2_+ +  
               \sqrt{6} \bigl( 1 - 3 q^2 \bigr) \lambda_q \cq_+ + 
               4 \cq_+ \bigl( \cp_+ + q \cp_- \bigr)                     
                       \bigl( q \cp_+ - \cp_- \bigr) \nonumber \\
            \quad +
               \Pi_- \bigl( \cp_+ - q \cp_- \bigr) +
               \Pi_+ \bigl( q \cp_+ + \cp_- \bigr) 
            \approx 0. 
         \end{eqnarray*}   
         If $q = \sigma/ \sqrt{3}$, the involution algorithm is closed when 
         $\varphi_q^{(5)}$ is computed, but it does not lead to any exact 
         solutions since the super-Hamiltonian is not compatible with the 
         constraints $\varphi_q^{(j)}$, with $j=1,\dots,5$. On the other hand,
         if $q \neq \sigma/ \sqrt{3}$, we have not been able to close the 
         algorithm. It turns out, however, that the involution algorithm in 
         this case is useless since the system under consideration does not 
         produce an integration case, as it can be shown by a local study of 
         its analytic structure.
\end{enumerate}

\subsubsection{Constraints $\Pi_{\pm} \approx 0$.}
\label{subsubsec:PiCons}

Consider now the extra constraints expressing that the canonical variables
$\Pi_{\pm}$ be equal to zero. Contrary to the involution of the previous 
constraints $\varphi_p$ and $\varphi_q$, the consistency algorithm here is 
trivial: it is unable to produce any new information since there are no 
secondary constraints. The constraints $\Pi_{\pm} \approx 0$ are first class; 
we can choose their corresponding Lagrange multipliers $\lambda_{\pm}$ to be 
zero. In that case the constraints remain weak equations to be imposed on the
physical states of the system. Hence the appropriate system of canonical 
equations is the system \eref{CanEq1a}--\eref{CanEq1b}, where we set 
$\Pi_{\pm}$ to zero. Beside equation \eref{CanEq1a}, the relevant equations are
thus 
\numparts
\begin{eqnarray}
   \dot{\cq}_+ &=&   \frac4{\sqrt{6}} 
                     \bigl( \cp_+ \cq_+ - \cp_- \cq_- \bigr), 
                     \label{PiEq1a} \\
   \dot{\cq}_- &=& - \frac4{\sqrt{6}} 
                     \bigl( \cp_- \cq_+ + \cp_+ \cq_- \bigr), 
                     \label{PiEq1b} \\
   \dot{\cp}_+ &=&   \frac2{\sqrt{6}} \bigl( \cp^2_- - \cp^2_+ \bigr) + \cq_+,
                     \label{PiEq1c} \\         
   \dot{\cp}_- &=&   \frac4{\sqrt{6}} \cp_+ \cp_-  + \cq_-. 
                     \label{PiEq1d}         
\end{eqnarray} 
\endnumparts
The super-Hamiltonian \eref{SupHam2} is now given by
\begin{equation} \label{SupHam5}
   \EuH_{\rmc} = - \frac2{\sqrt{6}} 
                      \Bigl[ 
                         \cq_+ \bigl( \cp^2_- - \cp^2_+ \bigr) +
                         2 \cp_+ \cp_- \cq_-    
                      \Bigr]
                 - \frac12 \bigl( \cq^2_+ + \cq^2_- \bigr).  
\end{equation}
\indent
The general solution of equations \eref{PiEq1a}--\eref{PiEq1d} is easy to 
produce under analytic form. We first solve equations \eref{PiEq1c} and 
\eref{PiEq1d} for $\cq_+$ and $\cq_-$ and write down second-order equations for
$\cp_+$ and $\cp_-$, namely $3 \ddot{\cp}_\pm = 4 \Sigma \cp_{\pm}$, with 
$\Sigma := \cp_+^2 + \cp_-^2$. Beside the super-Hamiltonian constraint, 
$\EuH_{\rmc} \approx 0$, this system possesses the first integral 
$\cp_+ \dot\cp_- - \cp_- \dot\cp_+ = \delta$, where $\delta$ denotes an 
arbitrary constant. Making use of the super-Hamiltonian constraint it is 
straightforward to produce a scalar second-order equation for $\Sigma$, namely
$\ddot\Sigma = 4 \Sigma^2$. We integrate this last equation so as to produce 
the solution $2 \Sigma = 3 \wp (t - t_0; 0, g_3)$, with arbitrary constants 
$t_0$ and $g_3 = \case{16}{9} \delta^2$. This reduces the system
$3 \ddot\cp_{\pm} = 4 \Sigma \cp_{\pm}$ to linear differential equations of 
the Lam\'{e} type, namely $\ddot\cp_\pm = 2 \wp (t - t_0; 0, g_3) \cp_\pm$, 
studied exhaustively in references \cite{ince} and \cite{whiwa} (if we adopt 
Ince's notations, our particular case of the Lam\'{e} equation is specified by 
$h = 0$ and $n = 1$). The general solution of the system \eref{PiEq1a}--%
\eref{PiEq1d} is indeed uniform. This confirms the single-valuedness of a 
possible integrability case detected in Subsection
\ref{subsec:IntegrableCases}, namely that the complete system with 
$\Pi_\pm = 0$ has the Painlev\'{e} property; see Subsection 
\ref{subsec:Painleve}. If we introduce $t_{\rmz}$ such that the transcendental
equation $\wp (t_{\rmz}) = 0$ is satisfied, \ie $t_{\rmz}$ is a zero of the
Weierstrass elliptic function, then the general solution to the system under
study is given by the fundamental set
\begin{equation} \label{LamGen}
   \cp_{\pm, 1} = \exp \bigl[ - t \zeta (t_{\rmz}) \bigr]
                  \frac{\sigma \bigl( t + t_{\rmz} \bigr)}{\sigma (t)}, 
   \quad
   \cp_{\pm, 2} = \exp \bigl[ + t \zeta (t_{\rmz}) \bigr]
                  \frac{\sigma (t - t_{\rmz})}{\sigma (t)},
\end{equation}
with Weierstrassian functions $\zeta (t)$ and $\sigma (t)$ defined by 
$\dot{\zeta}(t) + \wp (t) = 0$ and 
$\dot{\sigma}(t) - \sigma (t) \zeta (t) = 0$ respectively. The solutions given 
by the fundamental set \eref{LamGen} are distinct, provided $e_i \neq 0$, with 
$i = 1, 2, 3$, which is indeed the case here when $g_3 \neq 0$, since $e_i$, 
with $i = 1, 2, 3$, are defined by
\begin{equation*}
   e_1 + e_2 + e_3 = 0, \quad
   4 \bigl( e_2 e_3 + e_3 e_1 + e_1 e_2 \bigr) = - g_2 \equiv 0, \quad
   4 e_1 e_2 e_3 = g_3.
\end{equation*}
If $g_3 = 0 = \delta$, then $e_1 = e_2 = e_3 = 0$, and the solutions in the
fundamental set are not distinct. This case is better understood in the light
of its analytic structure. We shall see below that what is referred to as the
singularity family $f_{(3), \pm}$ in Subsection \ref{subsec:LocalProof} and 
\ref{app:painleve} becomes an exact two-parameter particular solution when 
$\Pi_\pm$ vanish. With $K$ constant, this is
\begin{equation} \label{LamF3}
\fl
\eqalign{%
    \cq_+ = - \frac{\sqrt{6}}2 \frac{1 \pm \sin 4K - 2 \sin^2 4K}
                                   {(t - t_0)^2}, \quad
   &\cq_- = - \frac{\sqrt{6}}2 \frac{\cos 4K \bigl( 1 \pm 2 \sin 4K \bigr)}
                                    {(t - t_0)^2}, \\
    \cp_+ = \pm \frac{\sqrt{6}}2 \frac{\sin 4K}{t - t_0}, \quad
   &\cp_- = + \frac{\sqrt{6}}2 \frac{\cos 4K}{t - t_0}.}
\end{equation}
Denoting $\chi:= t - t_0$ and integrating equations \eref{CanEq1a} we obtain
type I homogeneous metrics under the form
\begin{equation} \label{MetF3}
\fl
   \rmd s^2 = - \rmd \chi^2 
              + \chi^{\mp \sin 4K - \sqrt{3} \cos 4K} \rmd x^2
              + \chi^{\mp \sin 4K + \sqrt{3} \cos 4K} \rmd y^2
              + \chi^{\pm 2 \sin 4K} \rmd z^2. 
\end{equation}

\section{Analytic structure of the Bianchi type I system}
\label{sec:AnalyticStructure}

\subsection{The Painlev\'{e} strategy}
\label{subsec:Painleve}

The existence of the above solutions, whether particular solutions of the
general differential system or general solutions of specialised systems, tells
nothing about the integrability or non-integrability of the complete system 
and gives no information whatsoever about the mere accessibility of an exact 
and closed-form analytic expression of its general solution. This is due to 
the fact that the global involution algorithm of the extra constraints, as 
operated above, is not related with integrability and may even prove to be 
nonexhaustive. The integrability issue will be tackled through an invariant 
investigation method of intrinsic properties of the general solution. In 
particular, the result will not depend on specific choices of the metric,
within some well-defined equivalence class.
\nl
The approach advocated by Painlev\'{e} proceeds from the main observation,
nowadays frequently exemplified in various areas of theoretical physics, that 
all analytic solutions encountered are single-valued or multiple-valued
\emph{finite} expressions (possibly intricate) depending on a finite number of
\emph{functions}---solutions of linear equations, elliptic functions, and the 
six transcendental functions systematically extracted by Painlev\'{e} and 
Gambier. Therefore, the building blocks of this process are the functions, 
explicitly defined through their single-valuedness. This emphasises the 
relevance of the analytic structure of the solution, as uniformisability 
ensures adaptability to all possible sets of initial conditions. The next step
of this approach deals with integrability in some fundamental sense. Indeed, 
probing the analytic structure of a system requires a global, as opposed to 
local, integrability-related property, namely the \emph{Painlev\'{e} 
property}---we recall that a differential system possesses this property if,
and only if, its general solution is uniformisable or, equivalently, exhibits
no movable critical singularity. At this level, integrable systems are defined
in the sense of Painlev\'{e} as those systems that possess the Painlev\'{e}
property. The Painlev\'{e} property is invariant under arbitrary holomorphic
transformations of the independent variable and arbitrary homographic
transformations of dependent variables. In particular, our results are
invariant under arbitrary analytic reparameterisations of time. Moreover,
in keeping with the above observation, the requirement for global
single-valuedness ought to be relaxed so as to accommodate the definition
to integrability in the practical sense. Upon using broader classes of
transformations, the (unavoidable) analytic structure of the solution is
easily kept under control and either allows, or rules out, the possibility
to produce a mode of representation of the general integral.
\nl
Such an approach requires the analytic continuation of the solution in the
complex domain of the independent variable and computes generic behaviours of 
the solution in a vicinity of each movable singularity. The trivial part of 
the Painlev\'{e} method, known as the Painlev\'{e} test, produces necessary
but not sufficient conditions for a system to enjoy the Painlev\'{e} property
and requires local single-valuedness of the general solution in a vicinity of 
all possible families of movable singularities. We recall that movable 
essential singularities are difficult to handle, since one lacks methods to 
write down conditions under which they are indeed noncritical. For a tutorial
presentation of basic ideas and constructive algorithms aimed at the 
generation of integrability conditions, we refer to Conte \cite{conte} and 
Ramani, Grammaticos, and Bountis \cite{rgb}.

\subsection{Analytic proof of non-integrability}
\label{subsec:LocalProof}

The complete differential system \eref{CanEq1c}--\eref{CanEq1f} admits three 
distinct local leading behaviours in some vicinity of movable singularities, 
that is, three distinct singularity families, given in \ref{app:painleve}. The
first leading behaviour, denoted $f_{(1)}$, is valid in a vicinity of the 
movable point $t=t_1$ and exists only when $\Pi_+ \Pi_- \neq 0$. The existence
of the second singularity family, referred to as $f_{(2), \pm}$, valid in some
vicinity of the movable point $t=t_2$, requires 
$\Pi_- (2 \sqrt{3} \Pi_- \pm 3 \Pi_+) \neq 0$, respectively. The third and 
last leading behaviour, denoted $f_{(3), \pm}$, is valid in some vicinity of 
the movable point $t=t_3$ and is associated with another arbitrary constant 
parameter, denoted $K$ in \ref{app:painleve}. It exists only when 
$4 K \not \in \{ 0, \pi, - \pi, \frac{\pi}2, - \frac{\pi}2, \mp \frac{\pi}6, 
 \mp \frac{5 \pi}6 \}$. The super-Hamiltonian \eref{SupHam2} introduces no new 
restrictions since, at leading order in $\chi:= t - t_i$, for
all $i = 1, 2, 3$, it always vanishes in some vicinity of movable singularities 
around which leading behaviours $f_{(1)}$, $f_{(2), \pm}$, and $f_{(3), \pm}$ 
hold. Around each species of movable singularities one must inquire whether it
is possible to generate single-valued generic local representations of the 
general solution.
\nl
Such local expansions will be required to be both generic and single-valued.
As it is well known, both requirements are associated with Fuchsian indices of
each family. The indices have been computed for each leading behaviour and all 
preliminary conditions have proved to be fulfilled (for a thorough account an 
interested reader may consult \ref{app:painleve}). However, near both movable 
points $t_2$ and $t_3$ in $\mathbb{C}$, it is not possible to build single-%
valued generic local expansions, unless some constraints are imposed. 
Introducing movable logarithmic terms in these local series one may regain 
genericness---at the expense of losing the Painlev\'{e} property. The complete
proof of these claims can be found in \ref{app:painleve}.
\nl
Such multiple-valuedness, betrayed by local investigations, also pertains to
the general solution itself. This analytically proves that the system under
consideration is not integrable. Its general solution exhibits, in complex
time, an infinite number of logarithmic transcendental essential movable 
singularities; in other words, an analytic structure not compatible with
integrability in the practical sense: the quest for generic, exact and 
closed-form analytic expressions of the solution is hopeless. This result 
holds under space-time transformations within the equivalence class of the
Painlev\'{e} property; see Conte \cite{conte}. Yet, local information produced
by the Painlev\'{e} test may still be used in order to extract all particular
systems which may prove integrable. In the case under study, see Subsection
\ref{subsec:GloInv}, it turns out that all particular solutions are
meromorphic, but this is only a result of the application of the method.

\subsection{All integrable particular cases}
\label{subsec:IntegrableCases}

Even though this does not provide a proof but merely an indication of
uniformisability, the Painlev\'{e} test requires local single-valuedness 
around all possible species of movable singularities. We shall explicitly use
the results given in \ref{app:painleve}. Near the movable point $t=t_2$, the 
complete set of integrability conditions is satisfied if one imposes
\begin{equation} \label{CondF2}
   \Pi_- = \pm \sqrt{3} \Pi_+
\end{equation}
and requires the vanishing of the arbitrary constant parameter associated
with index $j = -3$, \ie $\tilde{\cq}_{+, -3}^{(1)} = 0$. Both conditions must
be required simultaneously. On the other hand, if one sets 
$2 \sqrt{3} \Pi_- = \mp 3 \Pi_+$ or $\Pi_- = 0$, the family $f_{(2), \pm}$ 
dies out and \emph{ipso facto} no restriction needs fulfilment.
\nl
Near the movable point $t = t_3$, single-valuedness may only be recovered in 
two particular cases. The first case deals with a local representation of the 
general solution of that specialised system obtained with $\Pi_\pm = 0$. In 
this case, local leading behaviours $f_{(1)}$ and $f_{(2), \pm}$ die out 
whilst, near $t = t_3$, a meromorphic local generic expansion is produced. 
Indeed, in this case, \S~\ref{subsubsec:PiCons} has already provided the
uniform general solution in closed analytic form. The second case deals with a
local representation of some particular solution of the complete differential
system and requires that the arbitrary parameter $K$ be such that
\begin{equation} \label{CondF3}
   \sin 4 K =     \frac{\Pi_+}{\sqrt{\Pi_+^2 + \Pi_-^2}}, \quad
   \cos 4 K = \pm \frac{\Pi_-}{\sqrt{\Pi_+^2 + \Pi_-^2}}.
\end{equation}

\section{Conformal relationship with Einstein spaces}
\label{sec:einspace}

A four-dimensional space-time $(\EuM,g)$ can be mapped onto an \einspace, 
under a conformal transformation, 
$\tilde{g}_{ab}=\rme^{2 \sigma (\bi{x})} g_{ab}$, if and only if there exists 
a smooth function $\sigma (\bi{x})$ that satisfies 
\begin{equation} \label{Eisen}
   L_{ab} = \nabla_a \sigma \nabla_b \sigma - \nabla_a \nabla_b \sigma 
            - \frac12 g_{ab} g^{mn} \nabla_m \sigma \nabla_n \sigma
            - \frac{\Lambda}{6} \rme^{2 \sigma} g_{ab},
\end{equation}
where the tensor $L_{ab}$ is defined by equation \eref{LTens} and 
$\Lambda:=\tilde{R}/4$ denotes the cosmological constant characterising the
\einspace $(\EuM,\tilde{g})$ \cite{eisen}. The first integrability condition 
of equation \eref{Eisen} is given by
\begin{equation} \label{IntCon1}
   \nabla_m C^m_{\ abc} + C^m_{\ abc} \nabla_m \sigma = 0.
\end{equation}
This is the necessary and sufficient condition for a space to be conformally
related to a ``$C$-space'' \cite{szeke}. The second integrability condition of
equation \eref{Eisen} is nothing else than equation \eref{Bach1}: fulfilment 
of the Bach equations is a necessary condition for a space to be conformally 
related to an \einspace. If considered separately the above integrability 
conditions of equation \eref{Eisen} are merely necessary conditions with 
regard to the conformal relationship with \einspaces. However, Kozameh \etal 
have proven that they constitute a set of sufficient conditions as well: 
A space-time $(\EuM,g)$ is conformally related to an \einspace 
$(\EuM,\tilde{g})$ if and only if equations \eref{Bach1} and \eref{IntCon1} 
are fulfilled \cite{kozam}.
\nl
Now let us specify the components of equation \eref{Eisen} for the Bianchi 
type I metric \eref{Misner}; we obtain the differential system
\numparts
\begin{eqnarray}
   2 \ddot{\sigma} - \dot{\sigma}^2 
      + 5 \bigl( \dot{\beta}_+^2 + \dot{\beta}_-^2 \bigr)
      - \frac{\Lambda}3 \rme^{2 \sigma} = 0, \\
   \dot{\sigma}^2 
      + 2 \dot{\sigma} \bigl( \dot{\beta}_+ + \sqrt{3} \dot{\beta}_- \bigr)
      + \ddot{\beta}_+ + \sqrt{3} \ddot{\beta}_-
      - \dot{\beta}_+^2 - \dot{\beta}_-^2
      - \frac{\Lambda}3 \rme^{2 \sigma} = 0, \\
   \dot{\sigma}^2 
      + 2 \dot{\sigma} \bigl( \dot{\beta}_+ - \sqrt{3} \dot{\beta}_- \bigr)
      - \ddot{\beta}_+ - \sqrt{3} \ddot{\beta}_- 
      - \dot{\beta}_+^2 - \dot{\beta}_-^2
      - \frac{\Lambda}3 \rme^{2 \sigma} = 0, \\
   \dot{\sigma}^2 
      - 4 \dot{\sigma} \dot{\beta}_+
      - 2 \ddot{\beta}_+
      - \dot{\beta}_+^2 - \dot{\beta}_-^2
      - \frac{\Lambda}3 \rme^{2 \sigma} = 0,
\end{eqnarray}
\endnumparts
which readily integrates to yield the necessary and sufficient condition for a
space of Bianchi type I to be conformally related to an \einspace, namely
\begin{equation} \label{EinSpa1}
   \cp_{\pm} = - \sqrt{6} \dot{\beta}_{\pm} 
             = - \sqrt{6} k_{\pm} \rme^{- 2 \sigma}, \qquad k_{\pm} \neq 0,
\end{equation}
and provides us with a differential equation that enables to determine 
explicitly the conformal factor $x:=\rme^{2 \sigma}$, \viz
\begin{equation} \label{EinSpa2}
   \dot{x}^2 = \frac43 \Lambda x^3 + 4 \bigl( k_+^2 + k_-^2 \bigr).
\end{equation}
This last equation can be written as the Weierstrass ODE in terms of the
Weierstrass elliptic function $\cw:=\wp (t - t_0; g_2, g_3)$, with invariants
$g_2 = 0$ and $g_3 = - \case49 \Lambda^2 \bigl( k_+^2 + k_-^2 \bigr)$.
Therefore, the conformal factor that brings a Bianchi type I space onto an
\einspace is given by
\begin{equation} \label{ConFac1}
   \rme^{2 \sigma (t)} = 3 \Lambda^{-1} \cw.
\end{equation}
Taking this last result into account we get from equation \eref{EinSpa1} the 
explicit form of the functions $\cp_\pm (t)$, that is
\begin{equation} \label{EinSpa3}
   \cp_{\pm} (t) = - \frac{\sqrt{6}}3 \Lambda k_{\pm} \cw^{-1}.
\end{equation}
So far we have not used the Bach equations \eref{BachEq00}--\eref{BachEq33} 
nor the equivalent canonical system \eref{CanEq1a}--\eref{CanEq1f}. If we do 
so, we see that the expression \eref{EinSpa3} does actually coincide with the 
solution \eref{BBSol} upon identifying 
\begin{equation} \label{Const1}
   p \equiv \frac{k_-}{k_+}, \qquad
   \Lambda \equiv \frac{k_+ \Pi_+ + k_- \Pi_-}{4 (k_+^2 + k_-^2)}.
\end{equation}
Thus we can rewrite type I homogeneous metrics \eref{MetPhip} under the 
equivalent form 
\begin{equation} \label{MetPhip2}
\eqalign{%
   \rmd s^2 = - \rmd t^2  
              &+ \exp \Biggl[ 
                         \pm \frac{2(k_+ + \sqrt{3} k_-)}
                                  {\sqrt{k_+^2 + k_-^2}} t \zeta(t_{\rmz})
                      \Biggr]
                 \Biggl[ 
                    \frac{\sigma (t-t_{\rmz})}{\sigma (t+t_{\rmz})}
                 \Biggr]^{\pm \frac{k_+ + \sqrt{3} k_-}
                         {\sqrt{k_+^2 + k_-^2}}} \rmd x^2 \\
              &+ \exp \Biggl[ 
                         \pm \frac{2(k_+ - \sqrt{3} k_-)}
                                  {\sqrt{k_+^2 + k_-^2}} t \zeta(t_{\rmz})
                      \Biggr]
                 \Biggl[ 
                    \frac{\sigma (t-t_{\rmz})}{\sigma (t+t_{\rmz})}
                 \Biggr]^{\pm \frac{k_+ - \sqrt{3} k_-}
                         {\sqrt{k_+^2 + k_-^2}}} \rmd y^2 \\
              &+ \exp \Biggl[ 
                         \mp \frac{4 k_+}{\sqrt{k_+^2 + k_-^2}} 
                         t \zeta(t_{\rmz})
                      \Biggr]
                 \Biggl[ 
                    \frac{\sigma (t-t_{\rmz})}{\sigma (t+t_{\rmz})}
                 \Biggr]^{\mp \frac{2 k_+}{\sqrt{k_+^2 + k_-^2}}} \rmd z^2.} 
\end{equation}
This is in agreement with the fact that imposing the constraint \eref{Phip0}%
---here, a direct consequence of equation \eref{EinSpa1}---on the Bach 
equations is equivalent to requiring condition \eref{IntCon1} to be fulfilled,
as it can be proved with the help of \textsc{Reduce}. In other words, assuming
that the ratio of variables $\cp_{\pm}$ is constant is a necessary condition
for a Bianchi type I space to be conformally related to an \einspace; in 
conformal gravity, it becomes a sufficient condition as well. Thus the only 
way to find out a solution to the Bach equations that is not conformally 
related to an \einspace is to relax the constraint \eref{Phip0}. In accordance
with our analysis of the preceding sections, we have indeed obtained the only 
closed-form analytical solution that cannot be mapped onto an \einspace, namely
the general solution \eref{LamGen} to the canonical system \eref{PiEq1a}--%
\eref{PiEq1d}, thereby confirming Schmidt's conjecture on the existence of such
solutions \cite{schmi}.
\nl
Now let us examine the particular case of a vanishing cosmological constant
$\Lambda$. \Eref{EinSpa2} becomes trivial and yields (up to an irrelevant 
constant of integration) 
\begin{equation} \label{ConFac2}
   \rme^{2 \sigma (t)} = t - t_0 =: \chi.
\end{equation}
Inserting this result into equation \eref{EinSpa1} and integrating we obtain 
type I homogeneous metrics under the form
\begin{equation} \label{LamNul1}
   \rmd s^2 = - \rmd \chi^2 
              + \chi^{2 (k_+ + \sqrt{3} k_-)} \rmd x^2
              + \chi^{2 (k_+ - \sqrt{3} k_-)} \rmd y^2
              + \chi^{- 4 k_+} \rmd z^2, 
\end{equation} 
which coincides with the exact 2-parameter particular solution \eref{MetF3} in
the case of zero $\Pi_{\pm}$, upon identifying 
$k_+ \equiv \mp \case12 \sin 4 K$ and $k_- \equiv - \case12 \cos 4 K$.
We then perform a conformal transformation with conformal factor 
\eref{ConFac2} and introduce the proper time $\tau:=\chi^{3/2}$; hence the 
metric \eref{LamNul1} that characterises the conformal \einspace with zero
cosmological constant is simply
\begin{equation} \label{LamNul2}
\fl
   \rmd s^2 = - \rmd \tau^2 
              + \tau^{\frac23 [2 (k_+ + \sqrt{3} k_-) + 1]} \rmd x^2
              + \tau^{\frac23 [2 (k_+ - \sqrt{3} k_-) + 1]} \rmd y^2
              + \tau^{\frac23 (1 - 4 k_+)} \rmd z^2. 
\end{equation} 
If we denote by $p_i$, with $i=1,2,3$, the successive exponents of $\tau$ in 
the spatial line element of the metric \eref{LamNul2}, we obtain
\begin{equation} \label{LamNul3}
   p_1   + p_2   + p_3 = 1, \qquad
   p_1^2 + p_2^2 + p_3^2 =  \frac13 \Bigl[ 1 + 8 (k_+^2 + k_-^2) \Bigr].
\end{equation} 
Computing the Bach equations for the metric \eref{LamNul1} we derive a 
constraint on the parameters $k_{\pm}$, namely $k_+^2 + k_-^2 = \case14$ (and 
thus $p_1^2 + p_2^2 + p_3^2 = 1$), that leads to the vacuum Bianchi type I
Kasner solution in \gr.

\ack

The authors are very grateful to H Caprasse and A Burnel for their suggestions
and comments. They thank Dr Schimming for hints to the literature. L Querella 
and C Scheen acknowledge financial support from the Belgian ``Fonds pour la 
formation \`{a} la recherche dans l'industrie et dans l'agriculture'' 
(\textsc{Fria} grants). This work was supported in part by contract No. 
\textsc{Arc} 94/99-178, ``Action de recherche concert\'{e}e de la 
Communaut\'{e} fran\c{c}aise''.

\appendix

\section{Bach tensor of the Bianchi type I space-time}
\label{app:bacheqn}

We have computed the components of the Bach tensor \eref{Bach2} for the 
Bianchi type I metric \eref{Misner}, with the \textsc{Excalc} package in 
\textsc{Reduce}. Up to irrelevant multiplicative factors, they are:
\begin{eqnarray}
\fl
   B_{00} = 2 \tdot{\beta}_- \dot{\beta}_- -
              \ddot{\beta}_-^2 -
           12 \dot{\beta}_-^4 -
           24 \dot{\beta}_-^2 \dot{\beta}_+^2 +
            2 \tdot{\beta}_+ \dot{\beta}_+ -
              \ddot{\beta}_+^2 -
           12 \dot{\beta}_+^4,
   \label{BachEq00} \\
\fl
   B_{11} = \sqrt{3} \bigl(    
                           \ddot{\ddot{\beta_-}} -
                        24 \ddot{\beta}_- \dot{\beta}_-^2 -
                         8 \ddot{\beta}_- \dot{\beta}_+^2 -
                        16 \dot{\beta}_- \ddot{\beta}_+ \dot{\beta}_+
                     \bigr) -  
            2 \tdot{\beta}_- \dot{\beta}_- +
              \ddot{\beta}_-^2 \nonumber \\ 
         - 16 \ddot{\beta}_- \dot{\beta}_- \dot{\beta}_+ + 
           12 \dot{\beta}_-^4 -  
            8 \dot{\beta}_-^2 \ddot{\beta}_+ +
           24 \dot{\beta}_-^2 \dot{\beta}_+^2 +
              \ddot{\ddot{\beta_+}} \nonumber \\
         -  2 \tdot{\beta}_+ \dot{\beta}_+ +
              \ddot{\beta}_+^2 -
           24 \ddot{\beta}_+ \dot{\beta}_+^2 +
           12 \dot{\beta}_+^4, 
              \label{BachEq11} \\
\fl 
   B_{22} = \sqrt{3} \bigl(    
                           \ddot{\ddot{\beta_-}} -
                        24 \ddot{\beta}_- \dot{\beta}_-^2 -
                         8 \ddot{\beta}_- \dot{\beta}_+^2 -
                        16 \dot{\beta}_- \ddot{\beta}_+ \dot{\beta}_+
                     \bigr) +  
              2 \tdot{\beta}_- \dot{\beta}_- -
                \ddot{\beta}_-^2 \nonumber \\
           + 16 \ddot{\beta}_- \dot{\beta}_- \dot{\beta}_+ -
             12 \dot{\beta}_-^4 +  
              8 \dot{\beta}_-^2 \ddot{\beta}_+ -
             24 \dot{\beta}_-^2 \dot{\beta}_+^2 -
                \ddot{\ddot{\beta_+}} \nonumber \\
           +  2 \tdot{\beta}_+ \dot{\beta}_+ - 
                \ddot{\beta}_+^2 +
             24 \ddot{\beta}_+ \dot{\beta}_+^2 -
             12 \dot{\beta}_+^4, 
              \label{BachEq22} \\
\fl
   B_{33} =  2 \tdot{\beta}_- \dot{\beta}_- -
               \ddot{\beta}_-^2 + 
            32 \ddot{\beta}_- \dot{\beta}_- \dot{\beta}_+ - 
            12 \dot{\beta}_-^4 - 
            16 \dot{\beta}_-^2 \ddot{\beta}_+ - 
            24 \dot{\beta}_-^2 \dot{\beta}_+^2 \nonumber \\
          +  2 \ddot{\ddot{\beta_+}} +
             2 \tdot{\beta}_+ \dot{\beta}_+ -    
               \ddot{\beta}_+^2 -
            48 \ddot{\beta}_+ \dot{\beta}_+^2 - 
            12 \dot{\beta}_+^4. 
              \label{BachEq33}
\end{eqnarray}

\section{Painlev\'{e} analysis}
\label{app:painleve}

Throughout this appendix we shall consider the auxiliary variables
\begin{eqnarray}
   \tilde{\cq}_\pm \equiv \cq_\pm^{a_\pm}
      &=& \sum_{n=0}^N \epsilon^n \sum_{j = n j_-}^{(n - N) j_- + j_+}
          \tilde{\cq}_{\pm, j}^{(n)} \, \chi^{j + q_\pm},
 \label{eqn:SeriesQpm} \\
   \tilde{\cp}_\pm \equiv \cp_\pm^{b_\pm}
      &=& \sum_{n=0}^N \epsilon^n \sum_{j = n j_-}^{(n - N) j_- + j_+}
          \tilde{\cp}_{\pm, j}^{(n)} \, \chi^{j + p_\pm}.
 \label{eqn:SeriesPpm}
\end{eqnarray}

\indent
Local analytic results presented here have been obtained through 
\textsc{Reduce} \cite{reduce} implementations of the Painlev\'{e} test
\cite{scheen}. The complete differential system possesses three distinct 
singularity families, referred to as $f_{(1)}$, $f_{(2), \pm}$, and 
$f_{(3), \pm}$. In the series \eref{eqn:SeriesQpm} and \eref{eqn:SeriesPpm} we 
have introduced:

\begin{itemize}
 \item the integer exponents $a_\pm$ and $b_\pm$, taking their values amongst 
       $\{-1,1\}$ and inducing homographic transformations of dependent 
       variables $\cq_\pm$ and $\cp_\pm$ (indeed, the Painlev\'{e} property is
       known to be preserved under such mappings);
 \item the order $N$ of the perturbative Painlev\'{e} test, taking its value
       amongst $\{0,\infty \}$;
 \item whenever $N \neq 0$, the small perturbative complex parameter
       $\epsilon$;
 \item the generically complex numbers $j_-$ and $j_+$, denoting respectively
       the lowest and highest Fuchsian indices---Fuchsian indices give the 
       orders at which arbitrary constant coefficients enter the double 
       expansion \eref{eqn:SeriesQpm}--\eref{eqn:SeriesPpm};
 \item $\tilde{\cq}_{\pm, j}^{(n)}$ and $\tilde{\cp}_{\pm, j}^{(n)}$, the 
       constant coefficients of the double expansion;
 \item the expansion variable $\chi$ of the inner Laurent series, given by 
       $t-t_1$, $t-t_2$, and $t-t_3$ in the cases $f_{(1)}$, $f_{(2), \pm}$, 
       and $f_{(3), \pm}$ respectively;
 \item the singularity orders, \ie leading powers $q_\pm$ and $p_\pm$;
 \item the nonvanishing coefficients $\tilde{\cq}_{\pm, 0}^{(0)}$ and 
       $\tilde{\cp}_{\pm, 0}^{(0)}$.
\end{itemize}

\indent
For each family, all aforementioned parameters are given by entries in 
table~\ref{tab:Families}, where \textsc{sf} stands for `singularity family' 
and with $s:=\sin 4K$ and $c:= \cos 4K$. In the case $f_{(3), \pm}$ the reason 
for the peculiar parameterisation of leading coefficients is that it produces
transparent integrability conditions; see below. The rightmost column provides
the set of Fuchsian indices.

\begin{table}
\caption{Families of movable singularities for the Bianchi type I space-time
         in conformal gravity.}
 \label{tab:Families}
 \begin{indented}
  \item[]
  \begin{tabular}{@{}lllllll}
   \br
   \textsc{sf} & $a_\pm, b_\pm$ & $N$ & $q_\pm, p_\pm$ &
   $\tilde{\cq}_{\pm, 0}^{(0)}, \tilde{\cp}_{\pm, 0}^{(0)}$
   & Indices \\
   \mr
   $f_{(1)}$ & $a_\pm = -1$ & $\infty$ & $q_\pm = -1$ &
   $\tilde{\cq}_{\pm, 0}^{(0)} = - \frac{\sqrt{6}}{\Pi_\pm}$ &
   $\{ -2, -2; -1, -1 \}$ \\
   & $b_\pm = -1$ & & $p_\pm = -2$ & $\tilde{\cp}_{\pm, 0}^{(0)}
   = - \frac{2 \sqrt{6}}{\Pi_\pm}$ & $(j_-, j_+) = (-2, -1)$ \\
   \mr
   $f_{(2), \pm}$ & $a_+ = -1$ & $\infty$ & $q_+ = -1$ &
   $\tilde{\cq}_{+, 0}^{(0)} = \pm \frac{3 \sqrt{2}}{\Pi_-}$ &
   $\{ -3; -1; 1; 3 \}$ \\
   & $a_- = -1$ & & $q_- = -1$ & $\tilde{\cq}_{-, 0}^{(0)}
   = \frac{9 \sqrt{2}}{2 \sqrt{3} \Pi_- \pm 3 \Pi_+}$ &
   $(j_-, j_+) = (-3, 3)$ \\
   & $b_+ = 1$ & & $p_+ = -1$ & $\tilde{\cp}_{+, 0}^{(0)}
   = - \frac{\sqrt{6}}{4}$ & \\
   & $b_- = 1$ & & $p_- = -1$ & $\tilde{\cp}_{-, 0}^{(0)}
   = \mp \frac{3 \sqrt{2}}{4}$ & \\
   \mr
   $f_{(3), \pm}$ & $a_+ = 1$ & $0$ & $q_+ = -2$ &
   $\tilde{\cq}_{+, 0}^{(0)} = - \frac{\sqrt{6}}{2}
   (1 \pm s - 2 s^2)$ & $\{ -1; 0; 3; 4 \}$ \\
   & $a_- = 1$ & & $q_- = -2$ & $\tilde{\cq}_{-, 0}^{(0)}
   = - \frac{\sqrt{6}}{2} c (1 \pm 2 s)$ & $(j_-, j_+) = (-1, 4)$ \\
   & $b_+ = 1$ & & $p_+ = -1$ & $\tilde{\cp}_{+, 0}^{(0)}
   = \pm \frac{\sqrt{6}}{2} s$ & \\
   & $b_- = 1$ & & $p_- = -1$ & $\tilde{\cp}_{-, 0}^{(0)}
   = \frac{\sqrt{6}}{2} c$ & \\
   \br
  \end{tabular}
 \end{indented}
\end{table}

\indent
Preliminary integrability conditions are fulfilled near all movable
singularities, located at $t=t_i$, for all $i=1,2,3$. Indeed, singularity 
orders, \ie leading powers $q_\pm$ and $p_\pm$, are negative integers; 
Fuchsian indices are integer numbers; maximal families are generated, thus 
producing local representations of the general solution, and rank conditions 
are satisfied by each set of Fuchsian indices---the multiplicity of each 
distinct Fuchsian index equals the number of arbitrary constant parameters to 
be introduced in the expansion at current Fuchsian index.
\nl
Perturbative approaches to the Painlev\'{e} test are required: the approach
due to Ablowitz, Ramani, and Segur \cite{ars,rgb}, in the case $f_{(3), \pm}$, 
and the Fuchsian perturbative approach devised by Conte, Fordy, and Pickering
\cite{conte,cfp,fp}, in the cases $f_{(1)}$ and $f_{(2), \pm}$. The
\emph{ansatz} previously introduced enables us to generate some integrability
conditions: for each value of $(n, j)$ such that $0 \leqslant n \leqslant N$
and $j$ is an index, an orthogonality condition must be fulfilled. Our results
may be summarised as follows.
\nl
For all $0 \leqslant n \leqslant 5$ and all $j \in \{-2,-1\}$, the
integrability conditions are fulfilled near $t=t_1$. The family $f_{(1)}$ does
seem to generate a single-valued local representation of the general solution,
but we recall that the Fuchsian perturbative method is not a bounded 
algorithm---\emph{stricto sensu} this is not an algorithm. However that may be,
this leading behaviour does not seem to be informative from the integrability
point of view.
\nl
On the contrary, dealing with the family $f_{(2), \pm}$, a local expansion near 
$t=t_2$ requires the presence of logarithmic terms in order to re-establish 
genericness. The $(n,j)=(0,3)$ integrability condition is violated, unless one
imposes the constraint 
\begin{equation} \label{eqn:CompleteCondF21}
   \frac{\pm \sqrt{3} \Pi_+ - \Pi_-}{\Pi_-} = 0.
\end{equation}
\noindent
Moreover, the constraint $\Pi_- = \pm \sqrt{3} \Pi_+$ does not preclude a 
subsequent incompatibility, encountered when $(n,j)=(1,3)$. This last 
incompatibility vanishes only if one freezes the arbitrary constant parameter 
associated with the index $j=-3$ so as to satisfy the relation
\begin{equation} \label{eqn:CompleteCondF22}
   \frac{4 \sqrt{6}}{81} \tilde{\cq}_{+, -3}^{(1)} \Pi_+^2 = 0
   \quad \Leftrightarrow \quad
   \tilde{\cq}_{+, -3}^{(1)} = 0.
\end{equation}
Since the multiplicity of index $j=-1$ is one, it is always possible upon a 
gauge choice to impose one scalar constraint amongst the two arbitrary 
parameters associated with that index $j=-1$, namely, $t_2$ and 
$\tilde{\cq}_{+, -1}^{(1)}$. We shall use this gauge freedom and set 
$\tilde{\cq}_{+, -1}^{(1)} = 0$. Under these circumstances, the Fuchsian 
perturbative method reduces to the test of Ablowitz, Ramani, and Segur---the 
algorithm stops.
\nl
We now consider the family $f_{(3), \pm}$ and produce integrability conditions
in a vicinity of the movable point $t=t_3$. Again, logarithmic terms are
required if one demands generic representations. The $(n,j)=(0,3)$
integrability condition is incompatible, unless the constraint
\begin{equation} \label{eqn:CompleteCondF31}
   \frac{\Pi_+ \cos 4 K \mp \Pi_- \sin 4 K}{\sin 4 K \cos 4 K} = 0
\end{equation}
is obeyed. The last occurrence of an index, namely $(n,j)=(0,4)$, gives no new
information---the index $j=4$ turns out to be compatible. Local single-%
valuedness may only be recovered in two particular cases:
\begin{enumerate}
   \item In the cases where both $\Pi_+$ and $\Pi_-$ vanish the general $4$-%
         parameter solution is locally single-valued near the movable point
         $t=t_3$. Indeed, when $\Pi_{\pm} = 0$ the only surviving family is 
         $f_{(3), \pm}$ and the weighted-homogeneous leading behaviour becomes
         an exact two-parameter meromorphic particular solution; see
         \S~\ref{subsubsec:PiCons}.
   \item When $\Pi_+ \neq 0$ and $\Pi_- \neq 0$, local single-valuedness may 
         only be recovered in the particular case 
         \begin{equation}  \label{eqn:CompleteCondF32}
            \sin 4 K := \frac{\Pi_+}{\sqrt{\Pi_+^2 + \Pi_-^2}}, \quad
            \cos 4 K := \pm \frac{\Pi_-}{\sqrt{\Pi_+^2 + \Pi_-^2}}.
         \end{equation}
\end{enumerate}
\indent
Integrability results are summarised in table~\ref{tab:Integrability}, which
gives, for each singularity family: all relevant existence conditions; 
nontrivial integrability conditions---\ie in\-com\-patibilities of conditions
for the system to possess the Painlev\'{e} property---; and all possibly 
integrable particular cases. The column providing the set of incompatibilities 
gives the value of $(n,j)$ where nontrivial conditions were met.

\begin{table}
 \caption{Local integrability conditions for the Bianchi type I space-time in 
          conformal gravity.}
 \label{tab:Integrability}
 \begin{indented}
  \item[]
  \begin{tabular}{@{}llll}
   \br
   \textsc{sf} & Existence conditions & Violations & Integrable cases \\
   \mr
   $f_{(1)}$ & $\Pi_+ \Pi_- \neq 0$ & $\emptyset$ & \\
   \mr
   $f_{(2), \pm}$ & $\Pi_- (2 \sqrt{3} \Pi_- \pm 3 \Pi_+) \neq 0$ &
   $(n, j) = (0, 3)$ & $\Pi_- = \pm \sqrt{3} \Pi_+$ \\
   & & and & and \\
   & & $(n, j) = (1, 3)$ & $\tilde{\cq}_{+, -3}^{(1)} = 0$ \\
   \mr
   $f_{(3), \pm}$ & $4 K \not \in \{ 0, \pi, - \pi, \frac{\pi}2, 
    - \frac{\pi}2, \mp \frac{\pi}6, \mp \frac{5 \pi}6 \}$ &
   $(n, j) = (0, 3)$ & $\Pi_+ = \Pi_- = 0$ \\
   & & & or \\
   & & & $\frac{\cos 4 K}{\sin 4 K} = \pm \frac{\Pi_-}{\Pi_+}$ \\
   \br
  \end{tabular}
 \end{indented}
\end{table}

\indent
Near the movable point $t=t_2$, the super-Hamiltonian constraint 
$\EuH_{\rmc} \approx 0$ requires $\tilde{\cq}_{+, 1}^{(0)} = 0$. Taking into
account all $f_{(2), \pm}$ integrability conditions one thus generates two-%
parameter meromorphic representations that begin like the truncated sums
\begin{eqnarray*}
   \tilde{\cq}_+ & \approx &
      \frac{\sqrt{6}}{\Pi_+} \chi^{-1} + \tilde{\cq}_{+, 3}^{(0)} \chi^2, \\
   \tilde{\cq}_- & \approx &
      \pm \frac{\sqrt{2}}{\Pi_+} \chi^{-1}
      \pm \Biggl( 
             \frac{2 \sqrt{2}}{9} - 
             \frac{\sqrt{3}}{9} \tilde{\cq}_{+, 3}^{(0)}
          \Biggr) \chi^2, \\
   \tilde{\cp}_+ & \approx &
      - \frac{\sqrt{6}}{4} \chi^{-1}
      + \Biggl( 
           \frac{\sqrt{6}}{6} - \frac{3}{4} \tilde{\cq}_{+, 3}^{(0)}
        \Biggr) \Pi_+ \chi^2, \\
   \tilde{\cp}_- & \approx &
      \mp \frac{3 \sqrt{2}}{4} \chi^{-1}
      \pm \frac{\sqrt{3}}{4} \tilde{\cq}_{+, 3}^{(0)} \Pi_+ \chi^2.
\end{eqnarray*}
Remarkably enough, if we set $\tilde{\cq}_{+, 3}^{(0)} := \frac{\sqrt{6}}6$ in
these expansions, we are brought back to manifolds with axial symmetry, since 
truncated-from-below Laurent series for $\tilde{\cp}_-$ and $\tilde{\cq}_+$
then identically reduce to the expansions for $\pm \sqrt{3} \tilde{\cp}_+$ and
$\pm \sqrt{3} \tilde{\cq}_-$ respectively.
\nl
Near the movable point $t=t_3$, the super-Hamiltonian constraint 
$\EuH_{\rmc} \approx 0$ requires $\tilde{\cp}_{-, 4}^{(0)} = 0$. Let us denote
$\sigma := \Pi_+^2 + \Pi_-^2$ and consider the family $f_{(3), \pm}$ in the 
integrable case such that $\sqrt{\sigma} \sin 4 K = \Pi_+$ and
$\sqrt{\sigma} \cos 4 K = \pm \Pi_-$. In such cases, the local Painlev\'{e}
test produces the two-parameter meromorphic series
\begin{eqnarray*}
   \tilde{\cq}_+ &\approx
      \mp \frac{\sqrt{6}}{2 \sigma}
         \biggl[ 
            \Pi_+ \sqrt{\sigma} \mp \bigl( \Pi_+^2 - \Pi_-^2 \bigr) 
         \biggr] \chi^{-2} \\
                 &\quad
      \pm \Biggl[ 
             \frac{\sqrt{6}}{12} \sqrt{\sigma}
             \pm \frac{\sqrt{6}}{12} \frac{\sigma}{\Pi_+}
             - 4 \tilde{\cp}_{-, 3}^{(0)} \Pi_-
               \frac{\sqrt{\sigma}}{\sigma}
             \mp 2 \tilde{\cp}_{-,3}^{(0)} \frac{\Pi_-}{\Pi_+}
          \Biggr] \chi, \\
   \tilde{\cq}_- &\approx
      \mp \frac{\sqrt{6}}{2 \sigma}
         \biggl[ \Pi_- \sqrt{\sigma} \pm 2 \Pi_+ \Pi_- \biggr] \chi^{-2} \\
                 &\quad
      \mp \Biggl[ 
             \frac{\sqrt{6}}{12} \frac{\Pi_-}{\Pi_+} \sqrt{\sigma}
             + 2 \tilde{\cp}_{-, 3}^{(0)} \frac{1}{\Pi_+}
                 \bigl( \Pi_+^2 - \Pi_-^2 \bigr)
                 \frac{\sqrt{\sigma}}{\sigma}
             \mp 2 \tilde{\cp}_{-, 3}^{(0)}
          \Biggr] \chi, \\
   \tilde{\cp}_+ &\approx
      \pm \frac{\sqrt{6}}{2} \frac{\Pi_+}{\sqrt{\sigma}} \chi^{-1} + 
      \Biggl[ 
         \frac{\sqrt{6}}{24} \frac{\sigma}{\Pi_+} - 
         \frac{\Pi_-}{\Pi_+} \tilde{\cp}_{-, 3}^{(0)}
      \Biggr] \chi^2, \\
   \tilde{\cp}_- &\approx
      \pm \frac{\sqrt{6}}{2} \frac{\Pi_-}{\sqrt{\sigma}} \chi^{-1} + 
      \tilde{\cp}_{-, 3}^{(0)} \chi^2.
\end{eqnarray*}
Again, we note that if one imposes 
$8 \tilde{\cp}_{-, 3}^{(0)} := \pm \sqrt{2} \Pi_+$ in these expansions, local 
representations of axially symmetric manifolds are produced.
\nl
To end this appendix, let us note that in cases such that both $\Pi_+$ and 
$\Pi_-$ vanish, a fourth singularity family, hereafter referred to as 
$f_{(4), \pm}$, may enrich the analytic structure of the system. In terms of 
the expansion variable $\chi := t - t_4$, this fourth local behaviour has 
leading terms
\begin{equation} \label{eqn:SingFamF4}
\fl
   \cq_+ \approx \pm \frac{\sqrt{3}}{3} \cq_{-, 0}^{(0)} \chi^2, \quad
   \cq_- \approx \cq_{-, 0}^{(0)} \chi^2, \quad
   \cp_+ \approx - \frac{\sqrt{6}}{4} \chi^{-1}, \quad
   \cp_- \approx \mp \frac{3 \sqrt{2}}{4} \chi^{-1},
\end{equation}
and four Fuchsian indices, namely $\{ -4; -1; 0; 3 \}$, for which the rank 
integrability condition is satisfied. However, when the super-Hamiltonian 
constraint $\EuH_{\rmc} \approx 0$ is to be satisfied, this implies the 
restriction $\mp 6 \sqrt{2} \cq_{-, 0}^{(0)} = 0$, and the family 
$f_{(4), \pm}$ dies out. Therefore, the new family enriches the analytic 
structure of the system only when $\EuH_{\rmc} \not \approx 0$.

\end{document}